\documentclass[12pt,titlepage]{article}

\usepackage{latexsym}
\usepackage{amsfonts}
\usepackage{amssymb}
\usepackage{epsfig}

\begin{document}


\begin{titlepage}

\title{Estimation of Collision Impact Parameter}
\author{M.~Savina, S.~Shmatov, P.~Zarubin \\
      \it Joint Institute for Nuclear Research, Dubna, Russia}
\date{}
\maketitle

\begin{abstract}
  We demonstrate that the nuclear collision geometry (i.e. impact parameter)
  can be determined with 1.5 fm accuracy in an event-by-event analysis by
  measuring the transverse energy flow in the pseudorapidity region
  $3 \le |\eta| \le 5$
  with a minimal dependence on collision dynamics details at the LHC energy
  scale.
  Using the HIJING model we have illustrated our calculation
  by a simulation of events of nucleus-nucleus interactions
  at the c.m.s energy from 1 up to 5.5 TeV per nucleon and various type
  of nuclei.
\end{abstract}

\end{titlepage}

\setcounter{page}{2}


The measurement of the collision impact parameter is a very
important practical problem of relativistic heavy ion physics. A part of
future experimental programmes on the LHC accelerator
will be devoted to relativistic heavy ion collisions \cite{CMS}.
Significant efforts will be focused on the
establishment of the fundamental laws and quite general rules of
nucleus-nucleus interactions, the discovery of a new state of
QCD matter, i.e. quark-gluon plasma (QGP), and the study of the properties of strongly
exited nuclear matter \cite{Gross81}. It is expected
that QGP might be produced in the {\it central} nucleus-nucleus collisions
at extremely high energy density $\epsilon_0 \sim$ 1 $GeV/fm^3$ (at the
LHC energy scale $\epsilon_0 \sim$ 1 $TeV/fm^3$).

On the other hand, for studies of diffractive
phenomena, properties of a coherent pomeron and collective nuclear
effects \cite{Bjorken92} it is necessary to select {\it peripheral} events with
a large collision impact parameter. Assuming that the collision impact parameter
is measured the experimentally observed effects can be compared with
theoreticall predictions of expected signals of a "new" physics:
parton energy losses in nuclear matter, monojet to dijet ratio,
quarkonia suppression, correlated jet and $W^{\pm}$,$Z^0$
production and so on.

In the present paper we demonstrate a method of
estimation of a collision impact parameter in the event-by-event
analysis.

As the collision impact parameter can not be measured in experiments
directly,
it would be necessary to find an experimentally measurable $b$-dependent
variable.
We suggest to use the total transverse energy $E_T$ (total energy $E$) produced
in the
one event of the nucleus-nucleus collision for the collision impact parameter estimation.

We will show that there is a correlation between
the transverse energy and the impact parameter, and will present our numerical
calculations performed on the basis of the HIJING model \cite{Wang91} at the RHIC and
LHC energy scales.
We argue that the measurement of $E_T$ produced in the forward direction (large
(pseudo)rapidity values) allows one to avoid possible uncertainties in the
$b$ determination.


The multiplicity and transverse energy production in the
nucleus-nucleus collisions in the ultrarelativistic energy domain
is considered as a combination of hard processes with $p_T \ge p_0$
and soft particle production.
A transverse energy flow produced in hard processes described by perturbative QCD
is associated in the main with minijet production, i.e. jets with $p_T \sim 4$
GeV  \cite{Eskola97}.

The average transverse energy carried by (mini)jets in the rapidity intervals $\Delta y$ is related
to the collision impact parameter $b$ by the formula:

\begin{equation}\label{eq:1}
 \langle E_T (b,\sqrt{s_{NN}},p_0,\Delta y) \rangle = T_{AA}(b)\sigma_{jet}(\sqrt{s_{NN}},p_0)_{\Delta y}
 \langle E_T^{\Delta y}\rangle,
\end{equation}
where $\langle E_T^{\Delta y} \rangle$ is the average transverse energy per a (mini)jet in $\Delta y$
interval, $\sigma_{jet}$ is the cross-section of minijet production in the parton model at the pp-level.
The differential distribution $d\sigma/dE_T$ is:

\begin{eqnarray}\label{eq:2}
     \frac{d\sigma_{jet}}{dE_T}(\sqrt{s_{NN}},p_0, \Delta y)=\frac{1}{2}K\int_{p_0^2, \Delta y}^{s/4}
     {dp_T^2 dy_1 dy_2} \sum_{i,j,k,l}{x_1 f_i(x_1,p_T^2) x_2 f_j(x_2,p_T^2)} \times \nonumber \\
     \qquad \lbrack \frac{d\hat {\sigma}^{ij \rightarrow kl}}{d \hat t}(\hat s, \hat t, \hat u)+
      \frac{d\hat {\sigma}^{ij\rightarrow kl}}{d \hat t}(\hat s, \hat u, \hat t) \rbrack
      \frac{\delta (E_T-p_T)}{1+\delta_{kl}},
\end{eqnarray}
where $p_0$ is the pQCD cut-off parameter, $x_1$ and $x_2$ are the fractional momenta of the initial
partons $i$ and $j$, and $y_1$ and $y_2$ are the rapidities of outgoing partons,
$d\hat \sigma^{ij\rightarrow kl}/d\hat t$ is the parton-parton scattering cross section. The summation runs
over all parton species and the factor $K \approx 2$ is
used to correct the lowest order pQCD rates for the effects of next-to-leading order terms.
The value of $\sigma_{jet}(\sqrt{s_{NN}},p_0)_{\Delta y}\langle E_{T}^{\Delta y}\rangle$
then is:

\begin{eqnarray}\label{eq:3}
     \sigma_{jet}(\sqrt{s_{NN}},p_0)_{\Delta y}\langle E_{T}^{\Delta y}\rangle =\frac{1}{2}K\int_{p_0^2, \Delta y}^{s/4}
     {dp_T^2 dy_1 dy_2} \sum_{i,j,k,l}{x_1 f_i(x_1,p_T^2) x_2 f_j(x_2,p_T^2)} \times \nonumber \\
     \qquad {\lbrack \frac{d\hat {\sigma}^{ij\rightarrow kl}}{d \hat t}(\hat s, \hat t, \hat u)+
      \frac{d\hat {\sigma}^{ij\rightarrow kl}}{d \hat t}(\hat s, \hat u, \hat t) \rbrack
      \frac{p_t}{1+\delta_{kl}}}.
\end{eqnarray}

Experimentally known effects of modification of quark and gluon structure functions by
nuclear medium called parton shadowing \cite{Shadowing} has not been included here.
To take into account the shadowing effect we should modify the formula (\ref{eq:3}) by
multiplying parton distributions by a corresponding correction term:

\begin{equation}
    f_i(x_1,p_T^2)f_j(x_2,p_T^2) \longrightarrow
    R_i^A(x_1,p_T^2)f_i(x_1,p_T^2)R_j^A(x_2,p_T^2)f_j(x_2,p_T^2),
\end{equation}
here the ratio $R_{i,j}^A \equiv f_{i,j/A}(x,p_T^2)/f_{i,j/N}(x,p_T^2)$, where $f_{i,j/N}(x,p_T^2)$ is
the parton structure function for a free nucleon and $f_{i,j/A}(x,p_T^2)$ is the corresponding
parton distribution in a proton inside the nucleus.

The nuclear density overlap function of two colliding nuclei at a given impact parameter
$T_{AA}(b)$ calculated in the assumption of the Wood-Saxon nuclear density distribution
$\rho_A(r)$:

\begin{equation}
   T_{AA}(|\vec{b}|)=\int_{}^{}{d^2\vec{r}\,T_A(\vec{r})T_A(\vec{b}-\vec{r})},
\end{equation}
where $\vec{r}$ is a 2-dimensional vector defining the interaction point.
The nuclear thickness function is

\begin{equation}
   T_{A}(|\vec{r}|)=\int_{}^{}{dz\rho_A(\sqrt{{|\vec{r}|}^2+z^2)}}.
\end{equation}

The expression (\ref{eq:1}) relating $E_T$ and $b$ includes a term arisen from hard
processes with $p_t \ge 2$ GeV only. For a more correct estimation of the collision
impact parameter we should take into account in addition
the part of a total transverse energy flow produced in soft interactions.

\begin{eqnarray}
  \langle E_T \rangle^{total} = \langle E_T \rangle^{jet}
  + \langle E_T \rangle ^{soft}, \nonumber \\[1em]
  \langle E_T \rangle^{soft} = T_{AA}(b)\sigma_{soft incl.}
  \langle E_T^{\Delta y} \rangle^{soft},
\end{eqnarray}
where $\langle E_T^{\Delta y} \rangle ^{soft}$ is the averaged
transverse energy per particle produced by soft interactions. But
soft processes can not be calculated by pQCD applications and we
should to use some phenomenological models \cite{Andersson87,
Nilson87} for the estimation of a soft part of the total energy
flow.



We think that a permissible pseudorapidity region for an the measuring energy is
limited by the following reason.
Main signatures of a possible QGP production are expected in the central
(pseudo)rapidity region. One of the
discussed features of such a state of nuclear matter is  energy losses of scattered partons in final
state interactions with a dense nuclear matter called jet quenching \cite{Wang97}. Among other
effects originated by jet quenching one may expect a significant modification in the distributions of
the transverse energy flow  and charged multiplicity $dE_T/d \eta$, $dE_T^{\gamma}/d \eta$,  and $dn_{ch}/d \eta$
\cite{Savina98}.


Indeed,  an indication is found for ultrarelativistic energy domain concerning the appearance of a wide bump
in the interval $-2 \le \eta \le 2$ over a pseudorapidity plateau of such distributions due to jet quenching.
Fig.1 demonstrates the evolution of the effect with collision impact parameter variation.
It is interesting to note that a secondary interaction effect such as a jet quenching  modifies only
the central rapidity part, while both pseudorapidity regions ($3 \le |\eta| \le 5$) remain
practically unchanged.
Therefore the large $\eta$ region can be used for the collision impact parameter estimation with minimal
dependence on possible signals of new physics in the central (pseudo)rapidity region.


The total cross-section of AA collisions has been calculated in the
framework of a HIJING hybrid model of nucleus-nucleus interactions \cite{Wang91},
where the cross-section of hard processes has been defined by the
formula (\ref{eq:3}). The contribution of the soft part of a produced
particle spectrum, i.e. small $p_T$-processes has been simulated
by using the FRITIOF and DPM models \cite{Andersson87, Nilson87}. In
these models hadrons are considered as relativistic strings exited
at hadron interactions. Calculations shows that for central PbPb
collisions at the LHC energy the inclusive cross-section of soft interactions at
the pp-level is equal to 57.0 mb while the inclusive cross-section of hard
processes is equal to 54.3 mb. The parton shadowing effect has been taken
into account.

However, it is obvious that the average transverse energy of
partons produced in hard processes $\langle E_T \rangle^{jet}$ (3
$\div$ 5 GeV for $|\eta|\leq$ 0.5) is larger than the average
transverse energy of soft partons $\langle E_T \rangle^{soft} \sim$
0.4 GeV. This fact reduces more strongly the relative contribution
of soft processes in the total transverse energy production. In the
LHC energy domain already 80 \% of the total transverse energy is
calculated by perturbative QCD application.
This allows one to
reduce ambiguities of $E_T$ calculations induced by the use of
phenomenology models to take into account small $p_T$-processes.

In the framework of the HIJING model we have simulated 10.000
events of $PbPb, NbNb, CaCa$ interactions at 5.5 TeV per nucleon.
The dependence of the total transverse energy produced in the
pseudorapidity interval $3 \le |\eta| \le 5$ on the collision
impact parameter is presented in fig.\ref{fig:2}. The variable
$E_T$ is connected with $b$. Some fluctuations in
nucleus-nucleus collision dynamics limit the impact parameter
estimation to a 1.5 fm precision.

The correlation of the same type is presented in  fig.\ref{fig:3}
with variation of a collision energy. It is shown that the increase of
the collision energy leads to an improvement of the accuracy
the impact parameter definition.


The correlation curve for the total energy flow is of the some shape.
The bulk of the energy produced in the forward direction run up to $10 \div 100$ $TeV$.
This allows one to measure the total energy with high accuracy and to reduce
experimental errors for the $b$ estimation.


We should remark that uncertainties are associated with the use of
various parton shadowing models and the sets of structure functions
lead to an ambiguity in $E_T-b$ correlation. In more detail the
influence of these aspects on the total (transverse) energy flow
is discussed in ref \cite{Zarubin2000}.

\newpage





\newpage

\begin{figure}[h,pt]
\begin{center}
\epsfig{figure=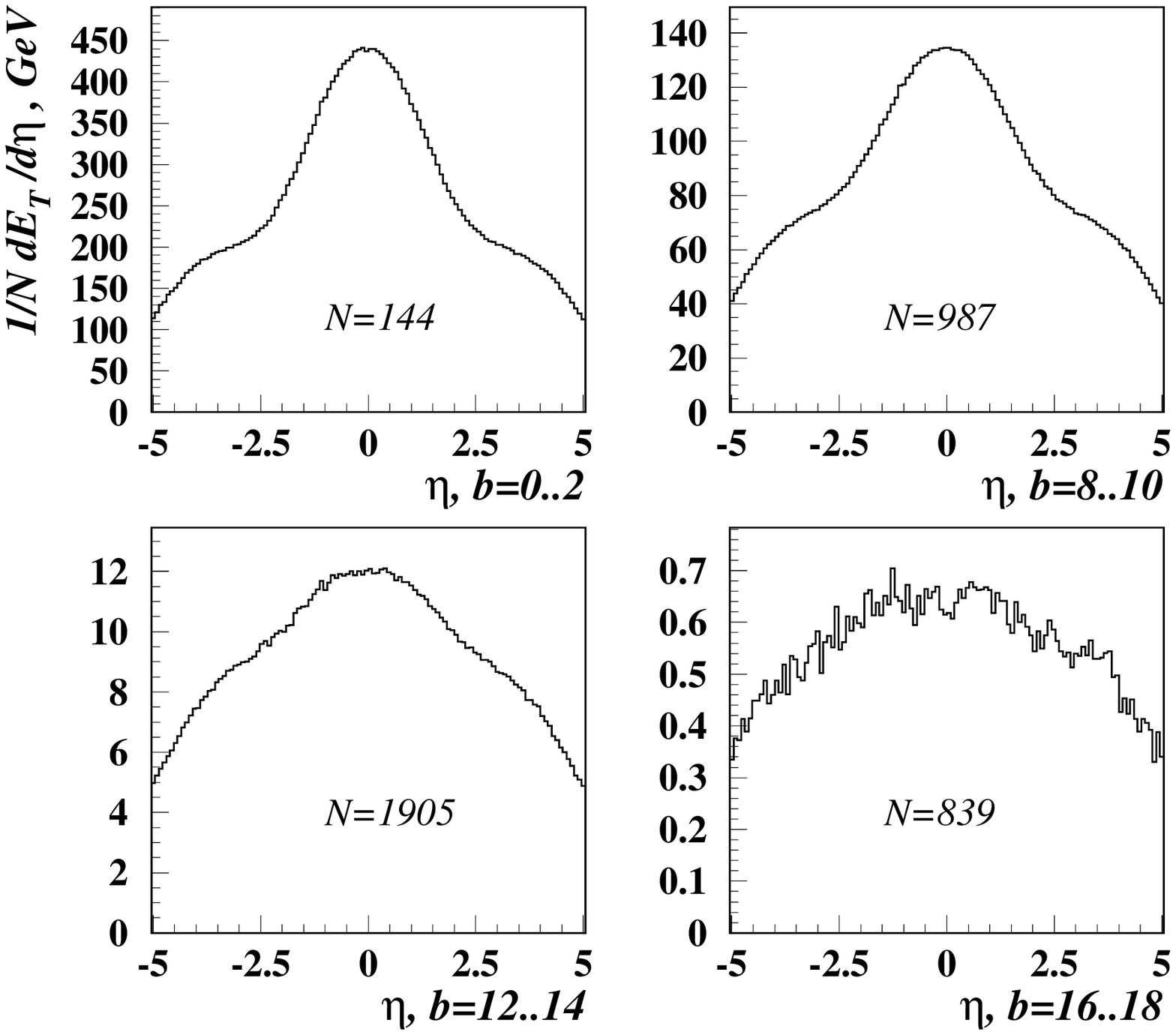, width=17cm}
\end{center}
\caption{Differential distribution of the total transverse energy $dE_T/d\eta$
(GeV) over pseudorapidity $\eta$ for 10000 minimum bias
PbPb collisions at $\sqrt{s_{NN}}$= 5.5 TeV/nucleon with various impact parameter.
Normalized per number of events.}
\label{fig:1}
\end{figure}

\begin{figure}[h,pt]
\begin{center}
\epsfig{figure=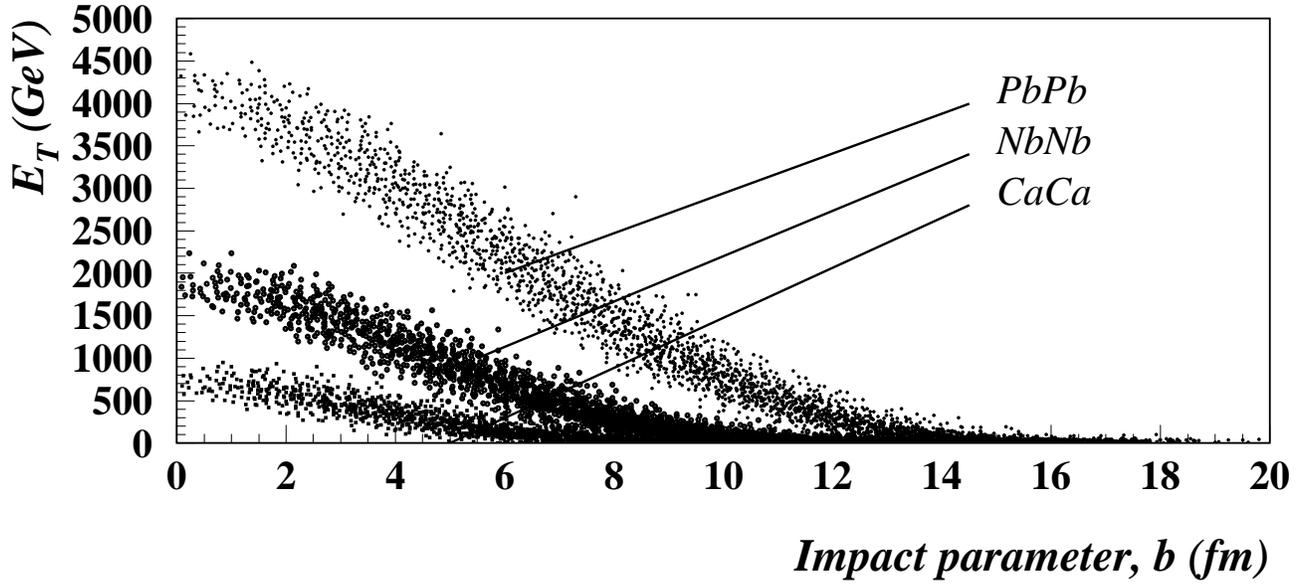, width=17 cm}
\end{center}
\caption{Correlation between the transverse energy flow per collision $E_T$ (GeV) in the
pseudorapidity direction (3$\le |\eta| \le $5) and collision impact parameter $b$ (fm).
From top to bottom: PbPb, NbNb, CaCa collisions at $\sqrt{s_{NN}}$=5.5 TeV/nucleon.}
\label{fig:2}
\end{figure}

\begin{figure}[h,pt]
\begin{center}
\epsfig{figure=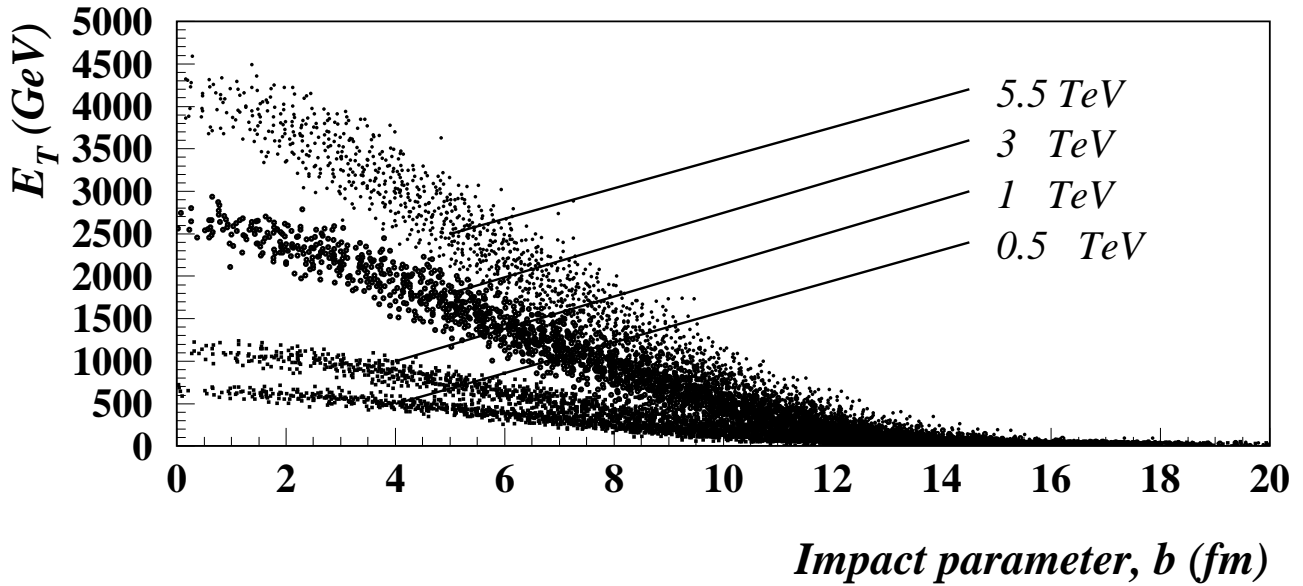, width=17cm}
\end{center}
\caption{Correlation between the transverse energy flow per collision $E_T$ (GeV) in the
pseudorapidity direction (3$\le |\eta| \le $5) and collision impact parameter
$b$ (fm) for PbPb collision at $\sqrt{s_{NN}}$=5.5, 3, 1, 0.5 TeV/nucleon.}
\label{fig:3}
\end{figure}

\end{document}